\pdfoutput=1

\documentclass[aps,pra,10pt,showpacs,superscriptaddress,longbibliography]{revtex4-1}

\usepackage{epsfig}
\usepackage{natbib}
\usepackage{amssymb}
\usepackage{soul}
\usepackage[breaklinks=true,colorlinks]{hyperref}

\begin{document}
\title{Quantum Superposition of Two Temperatures}

\author{Arun Kumar Pati}
\email{akpati@hri.res.in}
\affiliation{Quantum Information and Computation Group,
Harish-Chandra Research Institute, HBNI, Chhatnag Road, Jhunsi, Prayagraj 211 019, India}

\author{Avijit Misra}
\affiliation{Department of Chemical and Biological Physics,
Weizmann Institute of Science, Rehovot 7610001, Israel}
\affiliation{Optics and Quantum Information Group, The Institute of Mathematical Sciences,
HBNI, C.I.T Campus, Tharamani, Chennai 600 113, India}

\begin{abstract}
In the classical world, temperature is a measure of how hot or cold a physical object is. 
We never find a physical system which 
can be both hot and 
cold at the same time. Here, we show that for a quantum system, it is possible to have superposition of two temperatures which can lead to a situation that it can be found 
both in hot and cold state. We propose a physical mechanism for how to create a quantum state which is superposition of two temperatures. Furthermore, we define an operator 
for the inverse temperature
and show that the thermal state is, in fact, an eigenstate of this operator. The quantum state which represents superposition of two temperatures is not an eigenstate of
the inverse temperature operator. Our findings can have new applications in quantum thermodynamics, quantum nano scale devices and 
quantum statistical mechanics.
\end{abstract}
	
\date{\today}
\maketitle
\section{Introduction} 
Quantum superposition, which has been puzzling since its inception, lies at the heart of quantum theory. This superposition leads to quantum coherence \cite{CohRMP}, and exhibits
non-classical correlations \cite{EPR,EntRMP,Bera_2017,DisRMP} among distant parties that offer technological advances in communication \cite{Tele,Dense, Remote}, information processing \cite{QIP1,QIP2,QIP3}, computation \cite{Nielsen} and sensing \cite{QMetro1,QMetro2}.

While quantum theory is about a century old, thermodynamics \cite{callenbook} which stems from the quest of utilizing a energetic resource for various work efficiently has been well established 
over the centuries. This pragmatic branch of physics has witnessed several conceptual breakthroughs without violation of its principles \cite{Maxwell}. 

Only few decades ago people have started exploring thermodynamic behavior of objects at the atomic and nano-scales, in particular to explore what happens to thermodynamic
quantities at these scales where fluctuations cannot be ignored \cite{Jar1,Jar2}. Thus, stochastic thermodynamics \cite{Stocha1,Stocha2} and quantum thermodynamics 
\cite{Vinjanampathy_2016,R2,Kosloff_2013,David15} have emerged as active fields of research of late. Several intriguing
phenomena starting from heat flowing from cold to hot body \cite{ctoh}, to several second laws \cite{2ndlaws,Bera2017} have been established without contradicting the traditional
thermodynamics. Besides these 
foundational issues, designing quantum thermal machines \cite{OttoRonnie,QTM1} and exploring advantages of quantum resources like coherence and entanglement in
functioning of heat machines \cite{niedenzu18quantum,AREPL,CarnotPRE,Scully_2011,Raam,Huber,Deffner,FTE4} have also become active fields of study.

Classically the concept of temperature was originated as the average kinetic energy of an ideal gas \cite{callenbook}. Though the notion of temperature of a quantum particle is much more
intrigued, it can be visualized as the Lagrange multipliers that arises while maximizing the von Neumann entropy of a quantum particle with fixed average energy \cite{Jaynes1,Jaynes2}.
Thus, the temperature of the quantum particle can well be explained by the Gibbs distribution of it at some inverse temperature $\beta$. This lies at the heart of quantum 
thermometry \cite{Qthermo} where a quantum probe interacts with a heat bath and reaches the thermal distribution corresponding to the bath temperature. From this distribution of the probe, 
specifically some properties of the probe state that depends monotonically on this distribution, one can estimate the temperature of the probe. 
%

In the quantum world, physical objects can exist in a superposition of two
or more distinct states. Here, we ask can a quantum system be in a
superposition state characterized by two different temperatures? In classical systems, this is not possible. But in quantum mechanics the
possibility of quantum superposition of two temperatures is possible.
There is also experimental evidence that electrons carrying heat propagate
as two-dimensional quantum waves and the ripples in those waves are
predicted to lead to hot and cold spots \cite{Electemp}. This is in sharp
contrast to our everyday experience of temperature and heat flow. Even
though we do not intend to model the experiment, it is possible that there
are quantum superposition of two temperatures and the hot and the cold
spot could be due to quantum interference of these states.

While locally thermal states become important to inspect the role of quantum entanglement in storing work \cite{Workcor}, we inquire a fundamental question: Suppose we have two
physical systems at thermal distributions at two different temperatures. Is it possible to create a coherent quantum superposition of these two systems at two temperatures? 
We find that the answer to this question is yes. We show that if we take the purifications of the thermal states and perform Bell-state measurement on the ancillary systems, then that 
can create a quantum state which is superposition of two temperatures. Our protocol works in a probabilistic manner, but nevertheless brings out the main idea of our paper, i.e.,  
`making temperature quantum'.  We also show how the amplitudes of quantum superposition of two temperatures can display interference pattern
which is a key feature of quantum world. The interference pattern depends on the two different temperatures which may be hot and cold. Furthermore, we define an inverse temperature operator and show that the purified state of 
any thermal state is actually an eigenstate of this operator. That is consistent with our understanding that thermal states are in a definite temperature. We can also check that the superposition of two temperatures is not an eigenstate of the inverse temperature operator, thus hinting towards superposition of two temperatures.
Our findings could have a deep impact on the foundations
of quantum statistical physics and could have implications for quantum
nano technology, thermal machines at small scale and ultimately in quantum
technology.\\

  Before we present our main results, we would like to mention that if we have two thermal states at two
different temperatures, then we can imagine that there are two pure entangled
states corresponding to these two thermal states. Then, using the coherent quantum control or indefinite causal order \cite{causa1,causa2,causa3} one can create a 
superposition of these two pure entangled states. However, that will not corresponding to superposition of two temperatures in quantum world \cite{ved}.


\section{Superposition of temperatures} 
In this section, we provide a protocol where we can create a coherent superposition of quantum states with two distinct temperatures. We present our main idea for the case of qubits, but can be 
generalized for higher dimensions.

First, consider a simple case of two qubits with density operators $\rho_A = e^{-\beta_A H_A}/Z_A $ and $\rho_B = e^{-\beta_B H_B}/Z_B  $ which are in thermal states at 
inverse temperatures $\beta_A = 1/kT_A$ and $\beta_B = 1/kT_B$, respectively. Here $k$ is the Boltzmann constant and $T_A$, $T_B$ are temperatures of the thermal states $A$ and $B$.
Let the Hamiltonian of the two qubit systems be
\begin{eqnarray}
 H_A = E_0 |0\rangle \langle 0| +E_1 |1\rangle \langle 1|,\\
 H_B = E_0' |0\rangle \langle 0| + E_1' |1\rangle \langle 1|,
\end{eqnarray}
where $E_0$, $E_1$ are the eigenvalues of the Hamiltonian $H_A$ and $E_0'$, $E_1'$ are the eigenvalues of the Hamiltonian $H_B$.
The density matrix for the first qubit can be expressed as
\begin{equation}
 \rho_A= p_0|0\rangle \langle 0|+p_1|1\rangle \langle 1|,
\end{equation}
where $p_0=\frac{e^{-\beta_AE_0}}{Z_A}$, $p_1 = \frac{e^{-\beta_A E_1}}{Z_A} $ and $Z_A=e^{-\beta_AE_0}+e^{-\beta_AE_1}$ being the partition function.
Similarly, the density matrix for the second qubit can be expressed as 
\begin{equation}
\rho_B= f_0|0\rangle \langle 0|+f_1|1\rangle \langle 1|, 
\end{equation}
where $f_0=\frac{e^{-\beta_B E_0'}}{Z_B}$, $f_1 = \frac{e^{-\beta_B E_1'}}{Z_B}$ and $Z_B=e^{-\beta_B E_0'} + e^{-\beta_B E_1'}$ being the partition function. 
Now consider two purification of A and B such that $\rho_A = Tr_{A'}(|\Psi \rangle_{A'A} \langle \Psi | )$ and
$\rho_B = Tr_{B'}(|\Phi \rangle_{B'B} \langle \Phi | )$, where
\begin{eqnarray}
 |\Psi\rangle_{A'A} &=& \sqrt{p_0}|00\rangle + \sqrt{p_1}e^{i\phi} |11\rangle \\ \nonumber
 |\Phi\rangle_{B'B} &=& \sqrt{f_0}|00\rangle + \sqrt{f_1}|11\rangle 
\end{eqnarray}
Now consider the joint state of the composite system $AA'$ and $BB'$ as given by 
\begin{eqnarray}
 |\Psi\rangle_{A'A}|\Phi\rangle_{BB'} &=& ( \sqrt{p_0} |00\rangle_{A'A}+ \sqrt{p_1} e^{i\phi} |11\rangle_{A'A} ) \otimes (\sqrt{f_0}|00\rangle_{B'B} +  \sqrt{f_1} |11\rangle_{B'B})\\ \nonumber
 &=&\frac{1}{\sqrt{2}}[ |\Phi^+\rangle_{A'B'}(\sqrt{p_0 f_0} |00 \rangle_{AB} + \sqrt{p_1 f_1} e^{i\phi} |11\rangle_{AB} )\\ \nonumber
 &&+|\Phi^-\rangle_{A'B'} (\sqrt{p_0 f_0} |00\rangle_{AB} -  \sqrt{p_1 f_1} e^{i\phi} |11\rangle_{AB} ) \\ \nonumber
 &&+|\Psi^+\rangle_{A'B'}( \sqrt{p_0 f_1} |01\rangle_{AB}  + \sqrt{p_1 f_0} e^{i\phi} |10\rangle_{AB}) \\ \nonumber
 &&+|\Psi^-\rangle_{A'B``}(\sqrt{p_0 f_1} |01\rangle_{AB} - \sqrt{p_1 f_0} e^{i\phi} |10\rangle_{AB}) ],
\end{eqnarray}
where $|\Phi^{\pm}\rangle$ and $|\Psi^{\pm}\rangle$ are the well known Bell states. Now if we perform a Bell basis measurement on the system $A'B'$. If we find that
$|\Phi^{+}\rangle_{A'B'}$ clicks (for example), which has the probability $P_{\Phi^{+}}=\frac{1}{2}[p_0 f_0 + p_1 f_1]$ of occurrence, then the system $AB$ can be found 
in the following state
\begin{equation}
|\Psi(\Phi^{+}) \rangle_{AB}=\frac{1}{\sqrt{2}N}( \sqrt{p_0 f_0}  |00 \rangle_{AB} + \sqrt{p_1 f_1 } e^{i \phi}  |11\rangle_{AB}),
\end{equation}
where $N= \sqrt{p_0 f_0 + p_1 f_1 } $ is the normalization constant. Now, we can write the above state as
\begin{equation}
|\Psi(\Phi^{+}) \rangle_{AB}= \frac{1}{N\sqrt{2 Z_A Z_B}}(e^{-(\beta_A E_0 + \beta_B E_0')/2} |00\rangle_{AB} + e^{-(\beta_A E_1 + \beta_B E_1')/2}e^{i \phi}|11\rangle_{AB} ).\\
\end{equation}
In the above state, if we chose $E_0' =0$ and $E_1=0$, then we have a state  
\begin{equation}
\label{sot1}
|\Psi(\Phi^{+}) \rangle_{AB}= \frac{1}{N\sqrt{2 Z_A Z_B}}(e^{-\beta_A E_0/2} |00\rangle_{AB} + e^{- \beta_B E_1'/2 } e^{i\phi} |11\rangle_{AB}).\\
\end{equation}
The above state is a quantum superposition of two temperatures $\beta_A$ and $\beta_B$. The coherent oscillation between two amplitudes actually depends not on one temperature 
but on temperatures of two different baths which could be at $T_A$ and $T_B$. Also, this is not a superposition of two purified thermal states. This is genuine quantum superposition 
of two temperature state. If we view temperature as arising from the entanglement between the system and the bath, then the Bell measurement creates the entanglement between two 
independent qubits which are at two different temperatures. 

To see the coherent oscillation between two branches of the amplitudes, let us imagine that 
 we perform a controlled NOT gate on $AB$ and discard the system $B$. Then, we apply a Hadamard gate on the qubit $A$ and measure in the basis energy basis 
 $\{ |0 \rangle, | 1 \rangle  \}$. The probability of finding the system in these energy eigenbasis will reveal interefernce effect of two branches which are at two 
 different temperatures. For example, the probability of finding the qubit in the state $|0 \rangle $ is given by

\begin{equation}
\frac{1}{2}[1 +  \frac{1 }{ N^2 Z_A Z_B} e^{-(\beta_A E_0 +  \beta_B E_1')/2 } \cos \phi ].
\end{equation}

\begin{figure*}
  \centering
  \includegraphics[width=8cm]{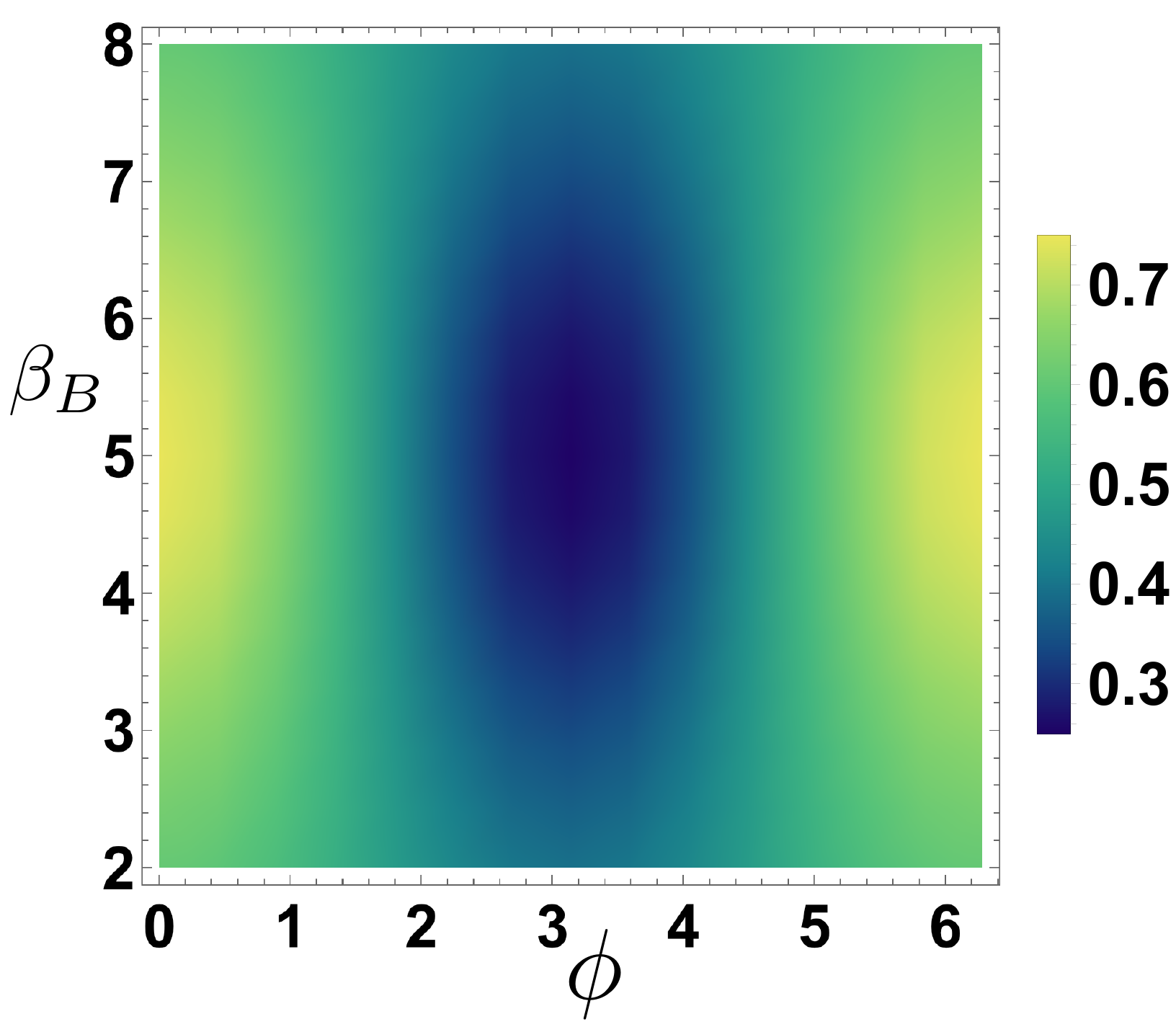}
  \caption{We plot the interference pattern of this superposition of two temperature states with $E_0=5$, $E_1'=1$ and $\beta_A=1$.}
  \label{Fig1}
\end{figure*}

The above expression displays interference which is a result of coherent superposition of two amplitudes at temperatures $T_A$ and $T_B$, respectively.  There is well defined
phase relationship between the amplitudes in the superposition.  We plot this interference pattern
in Fig. \ref{Fig1}

Let us explore what kind of superposition of two temperatures can be created if we get other outcomes.  If
$|\Phi^{-}\rangle_{A'B'}$ clicks, which also has the probability $P_{\Phi^{-}}=\frac{1}{2}[p_0 f_0 + p_1 f_1]$ of occurrence, then following the same procedure we get 
the state of the two qubits $AB$ as
\begin{equation}
|\Psi(\Phi^{-}) \rangle_{AB}= \frac{1}{N\sqrt{2 Z_A Z_B}}(e^{-\beta_A E_0/2} |0 0 \rangle_{AB} - e^{- \beta_B E_1'/2 } e^{i\phi} |1 1 \rangle_{AB})
\end{equation}
which differs from Eq. (\ref{sot1}) by a local unitary. For example, by applying $(I\otimes \sigma_z)$, we can obtain the same state as given in (9), i.e., 
a quantum state with superposition of two temperatures.

If
$|\Psi^{+}\rangle_{A'B'}$ clicks with probability $P_{\Psi^{+}}=\frac{1}{2}[p_0 f_1 + p_1 f_0]$,
then the system $AB$ can be found 
in the following state
\begin{equation}
|\Psi(\Psi^{+}) \rangle_{AB}= \frac{1}{N'\sqrt{2 Z_A Z_B}}(e^{-(\beta_A E_0 + \beta_B E_1')/2} |01\rangle_{AB} + e^{-(\beta_A E_1 + \beta_B E_0')/2}e^{i \phi}|10\rangle_{AB} ),\\
\end{equation}
where $N'= \sqrt{p_0 f_1 + p_1 f_0 }$.
Similarly, for $|\Psi^{-}\rangle_{A'B'}$  with probability $P_{\Psi^{-}}=\frac{1}{2}[p_0 f_1 + p_1 f_0]$, we get 
\begin{equation}
|\Psi(\Psi^{-}) \rangle_{AB}= \frac{1}{N'\sqrt{2 Z_A Z_B}}(e^{-(\beta_A E_0 + \beta_B E_1')/2} |01\rangle_{AB} - e^{-(\beta_A E_1 + \beta_B E_0')/2}e^{i \phi}|10\rangle_{AB} ).\\
\end{equation}
Clearly in the last two cases the choice $E_0' =0$ and $E_1=0$ does not work. Therefore, our protocol for creating superposition of two temperatures is probabilistic with success probability given by
\begin{equation}
 P_{suc}= p_0 f_0 + p_1 f_1.
\end{equation}

It may be noted that if we chose $E_1=0$ and $E_1'=0$, then we can create similar quantum superposition of two temperatures with total succuess probability $p_{suc} = [p_0 f_1 + p_1 f_0]$.
The quantum state which is a superposition of two  temperatures can have testable consequences. One can carry out an experiment to observe such
temperature superpositions. For example, if we create such a state and send through a Mach-Zehnder interferometer then  
we should be able to see an interference pattern
on the screen which will depend on the temperature of hot and cold bath.

\section{Inverse Temperature Operator}
 
If we think of making temperature as a quantum variable, then the natural question that comes to mind: Can there be a Hermitian operator 
for temperature? When a quantum state is in a thermal state, we know that it is in a thermal equilibrium and remains in a fixed temperature. Intuitively, one can then 
imagine that if such an operator exists for temperature, then the thermal state should be an eigenstate of this operator. 

Here, we would like to mention that after we defined the inverse temperature operator in 2017, we noticed that there is a proposal to define temperature as an 
operator \cite{Ghonge_2018}. 
However, our approach is different than what is given in the above paper.

In this section,  we define an inverse temperature operator and show that the purification of the thermal state is indeed an eigenstate of this operator. Let us consider a physical system $S$ which is in a thermal state at inverse temperature $\beta$ as given by
\begin{equation}
 \rho_S=  \frac{e^{-\beta H }} {Z}  = \sum_n \frac{e^{-\beta E_n}}{Z}|n\rangle \langle n|,
\end{equation}
where $H$ is the Hamiltonian of the system, $|n\rangle$ is the $n$th energy eigenstate of the Hamiltonian  with energy $E_n$ and $Z=\sum_ne^{-\beta E_n}$ is the partition 
function. Let us now consider a purification of $\rho_S$ as
\begin{equation}
 |\Psi\rangle_{SR} =\frac{1}{\sqrt{Z}}\sum_ne^{-\beta E_n/2}|n\rangle_S \otimes |n\rangle_R
\end{equation}
such that $\rho_S= Tr_{R}  (|\Psi\rangle_{SR}  \langle \Psi | )$ and $R$ denotes the bath degrees of freedom.

We define the (squared) inverse temperature operator ${\hat K} =  \sum_n (\frac{i\partial}{\partial E_n}\otimes\frac{-i\partial}{\partial E_n} ) $.  
This is a Hermitian operator that acts on ${\cal H}_S \otimes {\cal H}_R$. Moreover, the inverse temperature operator is a non-local operator on the purified Hilbert space. 
We will prove that the purification of any 
thermal state is an eigenstate of the (squared) inverse temperature operator with eigenvalue $c \beta^2$, where $c=16$, i.e., we have

\begin{equation}
{\hat K} |\Psi\rangle_{SR}  = \frac{\beta^2}{16}|\Psi\rangle_{SR}.
 \end{equation}

The action of the inverse temperature operator  ${\hat K}$ on the purified thermal state is given by
\begin{eqnarray}
 {\hat K} |\Psi\rangle_{SR} = &\sum_n& (\frac{i\partial}{\partial E_n}\otimes\frac{-i\partial}{\partial E_n} ) |\Psi\rangle_{SR} 
\\ \nonumber
 &=&\frac{1}{\sqrt{Z}}\sum_n\sum_m ( \frac{i\partial}{\partial E_n}\otimes\frac{-i\partial}{\partial E_n} ) e^{-\beta E_m/4}|m\rangle  \otimes e^{-\beta E_m/4}|m\rangle \\ \nonumber
 &=&\frac{1}{\sqrt{Z}}\sum_n\sum_m(\frac{-i\beta}{4})\delta_{nm}e^{-\beta E_n/4}|m\rangle\otimes(\frac{i\beta}{4})\delta_{nm}e^{-\beta E_n/4}|m\rangle  \\ \nonumber
 &=&\frac{\beta^2}{16}\frac{1}{\sqrt{Z}}\sum_ne^{-\beta E_n/2}|n\rangle \otimes |n\rangle= \frac{\beta^2}{16}|\Psi\rangle_{SR}.
\end{eqnarray}

 Therefore, $|\Psi\rangle_{SR}$ is an eigenstate of $\sum_n (\frac{i\partial}{\partial E_n}\otimes\frac{-i\partial}{\partial E_n})$ with eigenvalue $\frac{\beta^2}{16}$.
Thus, the  thermal state is, indeed, in a definite temperature state. However, the quantum superposition of two temperatures as given in (9) and (11) are not eigenstates of the 
(squared inverse) temperature operator. Thus, we cannot associate a definite temperature to the quantum superposition. Also, the 
individual qubits are not in a thermal state with definite temperature. More details of the inverse temperature operator and its applications to Thermodynamics uncertainty 
relations will be reported in future.

\section{Discussions and Conclusions} 

The notion of temperature, even though commonly used, it still raises deep questions when we apply thermodynamical concepts for quantum systems.
Recent progresses suggest that there can be thermalisation in isolated quantum systems which hints that a temperature can be assigned even to individual, pure quantum states. 
In classical physics, temperature of a physical system in thermal equilibrium can be found at a fixed temperature.  In this paper, 
we have shown that physical system can not only exists in a definite temperature state, but also in a superposition of two temperatures. In particular, we have given a protocol
how to create superposition of qubits  where the spin-up and spin-down states can be at two different temperatures. This coherent superposition of pure state
at two different temperatures can lead to quantum interference where interference pattern will depend on temperature of two baths. 
We have also defined a  (squared inverse) temperature operator and shown that the purification of the thermal state is an eigenstate of this operator.
This is in concurrence  with our  understanding that thermal state is in a definite temperature. The quantum superposition of two temperatures state is not an eigenstate 
of the temperature operator. We hope that our proposal for the superposition of two temperatures 
can be tested in experiment involving quantum interferometry.

Finally, it may be noted that if one uses two thermal channels and applies the coherent control of channels for the system along with a control qubit, 
one can create superposition of two purified thermal states. But we believe that it is not a 
superposition of two temperatures. We hope that our protocol will find useful applications in quantum thermodynamics, quantum nano scale devices and 
quantum statistical mechanics.

\vskip 1cm
\noindent
{\it Note added:-} After completion of our work in 2017, we came to know about the preprint \cite{Costa} which has addressed similar question, but the results are completely different.

\acknowledgments
AKP acknowledges local hospitality at the Institute of Mathematical Science, Chennai, India during his visit in Dec 2017 where this work was carried out.
AM acknowledges DAE postdoctoral fellowship at the Institute of Mathematical Sciences, Chennai, India.

\bibliography{SOT}
\end{document}